\documentclass[letterpaper,10pt]{article}
\usepackage{osameet2}

\usepackage{amsmath,amssymb}

\usepackage{graphicx}
\usepackage{caption}
\usepackage[textsize=scriptsize]{todonotes}

\usepackage{wrapfig}

\newcommand*{\vv}[1]{\vec{\mkern0mu#1}}

\usepackage[pdftex,colorlinks=true,bookmarks=false,citecolor=blue,urlcolor=blue]{hyperref} 

\begin{document}

\title{Concept and Experimental Demonstration of Optical IM/DD End-to-End System Optimization using a Generative Model\vspace*{-2ex}}

\author{Boris Karanov\textsuperscript{1,2}, Mathieu~Chagnon\textsuperscript{2}, Vahid~Aref\textsuperscript{2}, Domani\c{c} Lavery\textsuperscript{1}, Polina Bayvel\textsuperscript{1}, and Laurent Schmalen\textsuperscript{3}}
\address{\textsuperscript{1}Optical Networks Group, Dept. of Electronic \& Electrical Engineering, UCL, WC1E 7JE London, U.K. \\ \textsuperscript{2}Nokia Bell Labs, 70435 Stuttgart, Germany \\ \textsuperscript{3}Communications Engineering Lab, Karlsruhe Institute of Technology, 76131 Karlsruhe, Germany}
\email{boris.karanov.16@ucl.ac.uk}

\begin{abstract}
We perform an experimental end-to-end transceiver optimization via deep learning using a generative adversarial network to approximate the test-bed channel. Previously, optimization was only possible through a prior assumption of an explicit simplified channel model. \vspace*{-0.3ex}\mbox{$\copyright$ 2020 The Authors} \vspace*{-0.3ex}
\end{abstract}

\ocis{(060.4510) Optical communications, (060.2330) Fiber optics communications, (060.4080) Modulation\vspace*{-4ex}}

\section{Introduction}\vspace*{-1ex}
The optimization of a complete communication system via deep learning has attracted great interest since the introduction of the idea in~\cite{O'Shea_1}. The application of such end-to-end neural network-based \emph{autoencoders} is of particular importance in communication scenarios where the optimum transmitter-receiver pair is not known, or is computationally prohibitive to implement. An example is optical fibre communication based on intensity modulation and direct detection~(IM/DD), where the joint effects of chromatic dispersion, introducing intersymbol interference (ISI), and square-law detection render the channel nonlinear with memory. The absence of optimal, computationally feasible, processing algorithms for IM/DD systems prompted the introduction of end-to-end deep learning in~\cite{Karanov_1}, where it was shown experimentally that a simple feed-forward neural network~(FFNN) scheme can outperform pulse-amplitude modulation~(PAM) systems with simple feed-forward linear equalizers. Moreover, sequence processing by a recurrent neural network~(RNN) was used in~\cite{Karanov_2} to tailor the autoencoder to the dispersive channel properties. This led to an improved system performance with lower computational complexity than PAM transmission with receivers using nonlinear processing~\cite{Karanov_2} or maximum likelihood sequence detection~\cite{Karanov_3}.

The framework of learnable communication systems considers the channel as a part of the deep neural network. The application of autoencoders has thus been limited to scenarios where the channel model is known and differentiable. In addition, when applied to an actual system, a transceiver representation learned on a prior assumption of a specific model can result in sub-optimal performance~\cite[Sec. IV]{Doerner}~\cite[Sec. VI-C]{Karanov_1}. Fine-tuning the receiver part of the autoencoder with collected data from measurements was used in~\cite{Doerner}. However, optimizing the transmitter without the explicit knowledge of the underlying channel model remains an open problem. To circumvent using a model, a reinforcement learning method was proposed and verified for simple memoryless channels\cite{Aoudia}. Alternatively, a generative model of the channel can be obtained via a generative adversarial network~(GAN) and used for gradient backpropagation
\begin{wrapfigure}[10]{r}{0.63\textwidth}\vspace*{-3.75ex}
\centering
\includegraphics[width=0.63\textwidth, keepaspectratio = true]{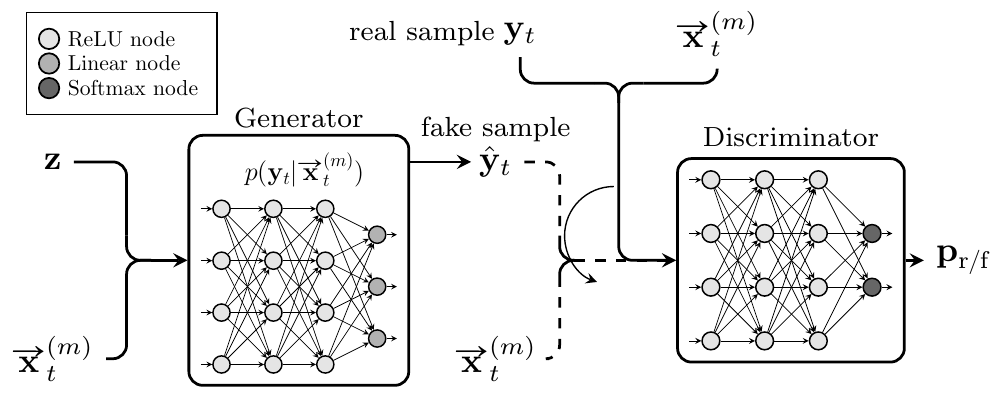}
\vspace*{-5ex}
\caption{Schematic of the conditional GAN structure for training of the generative model.}
  \label{fig:GAN_schematic}
\end{wrapfigure} 
during end-to-end optimization~\cite{O'Shea_2,Ye,Smith}.

We employ a GAN to acquire a model for our experimental IM/DD test-bed. We perform iterative steps of training the generative model on experimental data and use it to optimize the transceiver. To the best of our knowledge, this is the first end-to-end optimization of an optical communication system via deep learning based on measured data.\vspace*{-1ex}

\section{Generative Adversarial Network Design for the Optical Autoencoder}\label{sec:GAN}\vspace*{-1ex}
Generative adversarial networks are an effective tool for the learning of generative models~\cite{Goodfellow}. A GAN typically employs two artificial neural networks (ANNs) with competing objective functions, which are trained iteratively. The schematic of the GAN we used is shown in Fig.~\ref{fig:GAN_schematic}. The generator network aims at translating its input into a (possibly high-dimensional) output sample, mimicking some ground truth distribution of the data. The discriminator
\begin{wraptable}[13]{l}{0.18\textwidth}\vspace*{-1.5ex}
\caption{Generator (bold) and discriminator ANNs.}\label{tab_1}
\begin{tabular}{c|c}
    \hline
     Layer &  Outputs\\\hline
     \textbf{ReLU} & $30n$\\
     \textbf{ReLU} & $20n$\\
     \textbf{ReLU} & $13n$\\
     \textbf{ReLU} & $8n$\\
     \textbf{ReLU} & $5n$\\
     \textbf{Linear} & $n$\\\hline
     ReLU & $16n$\\
     ReLU & $10n$\\
     ReLU & $6n$\\
     Softmax & $2$\\\hline
\end{tabular}
\vspace*{-1ex}
  \label{tab:GAN_param}
\end{wraptable}
acts as a binary classifier between real (ground truth) and fake (generated) samples. Related to communication systems, GANs can be used to train a generative model which mimics the function of the channel by approximating its conditional probability distribution. In the framework of autoencoder design, the model can be used in the end-to-end system optimization. In this work we employ a simple FFNN autoencoder. The transmitter maps the input messages $s_{t}\in\{1,\ldots, S\}$, each of which carrying $\log_2(S)$ bits, into blocks (symbols) of $n$ samples, which we denote as $\mathbf{x}_t\in{\mathbb{R}}^{1\times n}$. After propagation through the channel, the symbols are fed to the receiver as $\mathbf{y}_t\in{\mathbb{R}}^{1\times n}$. Since ISI in optical IM/DD stems from both preceding and succeeding symbols, the goal of the generative model is to approximate $p(\mathbf{y}_t|\vv{\mathbf{x}}_t^{(m)})$, where $m$, an odd integer, is the modeled symbol memory and $\vv{\mathbf{x}}_t^{(m)} := \begin{pmatrix}\mathbf{x}_{t-(m-1)/2},\ldots,\mathbf{x}_{t+(m-1)/2}\end{pmatrix}\in{\mathbb{R}}^{1\times m\cdot n}$. To mimic the probability distribution, the generator takes as an input the concatenation of a vector of uniformly-distributed random samples $\mathbf{z} \sim \mathcal{U}_{1\times m\cdot n}^{(0,1)}$ with $\vv{\mathbf{x}}_t^{(m)}$. It transforms it into the fake symbol $\hat{\mathbf{y}}_t=G_{\boldsymbol{\theta}}\left((\mathbf{z},\vv{\mathbf{x}}_t^{(m)})\right)$, where $\hat{\mathbf{y}}_t\in{\mathbb{R}}^{1\times n}$. The discriminator is first fed with the real $\mathbf{y}_t$ and then the fake symbol $\hat{\mathbf{y}}_t$, both conditioned on $\vv{\mathbf{x}}_t^{(m)}$, producing the output probability vectors $\mathbf{p}_{\textnormal{r}}=D_{\boldsymbol{\phi}}\left((\mathbf{y}_t,\vv{\mathbf{x}}_t^{(m)})\right)$ or $\mathbf{p}_{\textnormal{f}}=D_{\boldsymbol{\phi}}\left((\hat{\mathbf{y}}_t,\vv{\mathbf{x}}_t^{(m)})\right)$, respectively, where $\mathbf{p}_{\textnormal{r}/\textnormal{f}}\in{\mathbb{R}}^{1\times 2}$. Table~\ref{tab_1} shows the layers of the two ANNs, trained in a supervised manner. The sets of parameters $\boldsymbol{\phi}$ (discriminator) and $\boldsymbol{\theta}$ (generator) are iteratively updated via stochastic gradient descent (SGD), using the Adam algorithm, aimed at minimizing the losses\vspace*{-1.5ex}
\begin{equation}\label{eq:D_G_loss}
    \overline{\mathcal{L}_D}(\boldsymbol{\phi})=\frac{1}{B}\sum\limits_{i=1}^{B}[\ell(\mathbf{l}_{\textnormal{r}},\mathbf{p}_{\textnormal{r},i})+\ell(\mathbf{l}_{\textnormal{f}},\mathbf{p}_{\textnormal{f},i})], \qquad \overline{\mathcal{L}_G}(\boldsymbol{\theta})=\frac{1}{B}\sum\limits_{i=1}^{B}\ell(\mathbf{l}_{\textnormal{r}},\mathbf{p}_{\textnormal{f},i}),\vspace*{-1.5ex}
\end{equation}
over a batch of $B\!=\!10^{3}$ elements from the training set, where $\mathbf{l}_{\textnormal{r}}\!:=\!(0,1)$ and $\mathbf{l}_{\textnormal{f}}\!:=\!(1,0)$ are the labels for real and fake symbols, respectively, and $\ell(\mathbf{x},\mathbf{y})\!=\!-\sum_{i}x_i\log(y_i)$ is the cross entropy. Note that the objective of the discriminator is to classify $\mathbf{y}_t$ and $\hat{\mathbf{y}}_t$ correctly. By minimizing~$\overline{\mathcal{L}_G}$, the generator learns representations $\hat{\mathbf{y}}_t$ that are statistically similar to $\mathbf{y}_t$. We train the GAN over $10^{4}$ steps, each of 4 consecutive discriminator updates with the learning rate of $10^{-3}$, followed by a generator update with the rate gradually reduced every 200 steps from $5\!\cdot\!10^{-4}$ to $10^{-5}$.\vspace*{-1ex}

\section{End-to-End Optimization Algorithm and Experimental Performance Results}\vspace*{-1ex}

\begin{figure}[b!]\vspace*{-4ex}
\centering
\includegraphics[width=\textwidth, keepaspectratio=true]{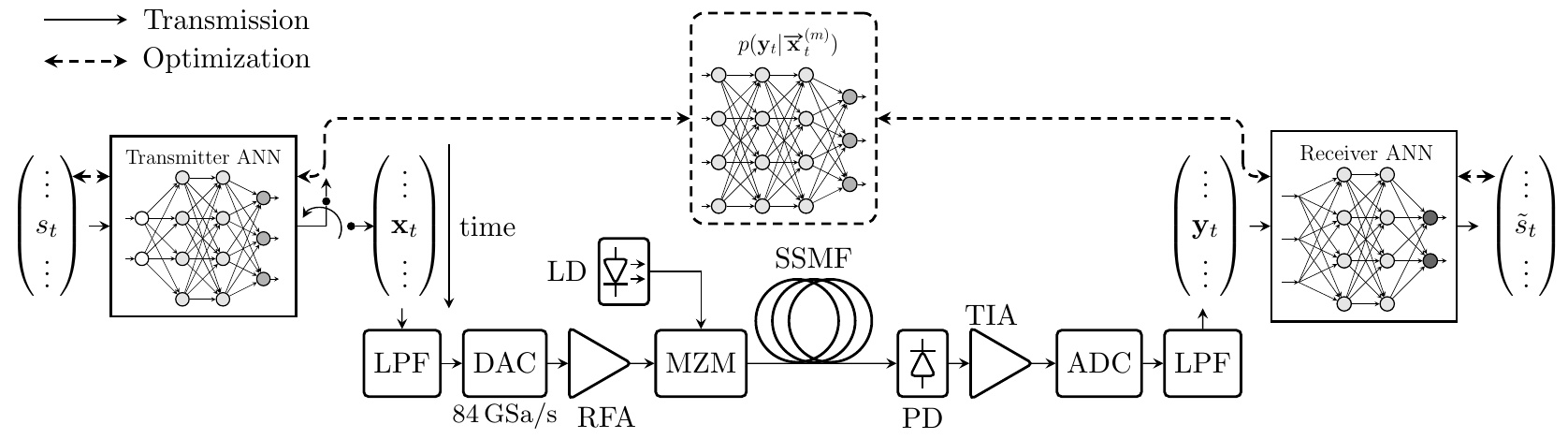}
\vspace*{-5.5ex}
\caption{Schematic of the IM/DD experiment, showing the optimization via the generative channel model. LPF: low-pass filter, DAC/ADC: digital-to-analog/analog-to-digital converter, MZM: Mach-Zehnder modulator, SSMF: standard single mode fibre, TIA: trans-impedance amplifier.}
\label{fig:Exp_setup}
\vspace*{-3ex}
\end{figure}
\begin{figure}[t!]\vspace*{-2ex}
\centering
\includegraphics[width=1\textwidth, keepaspectratio=true]{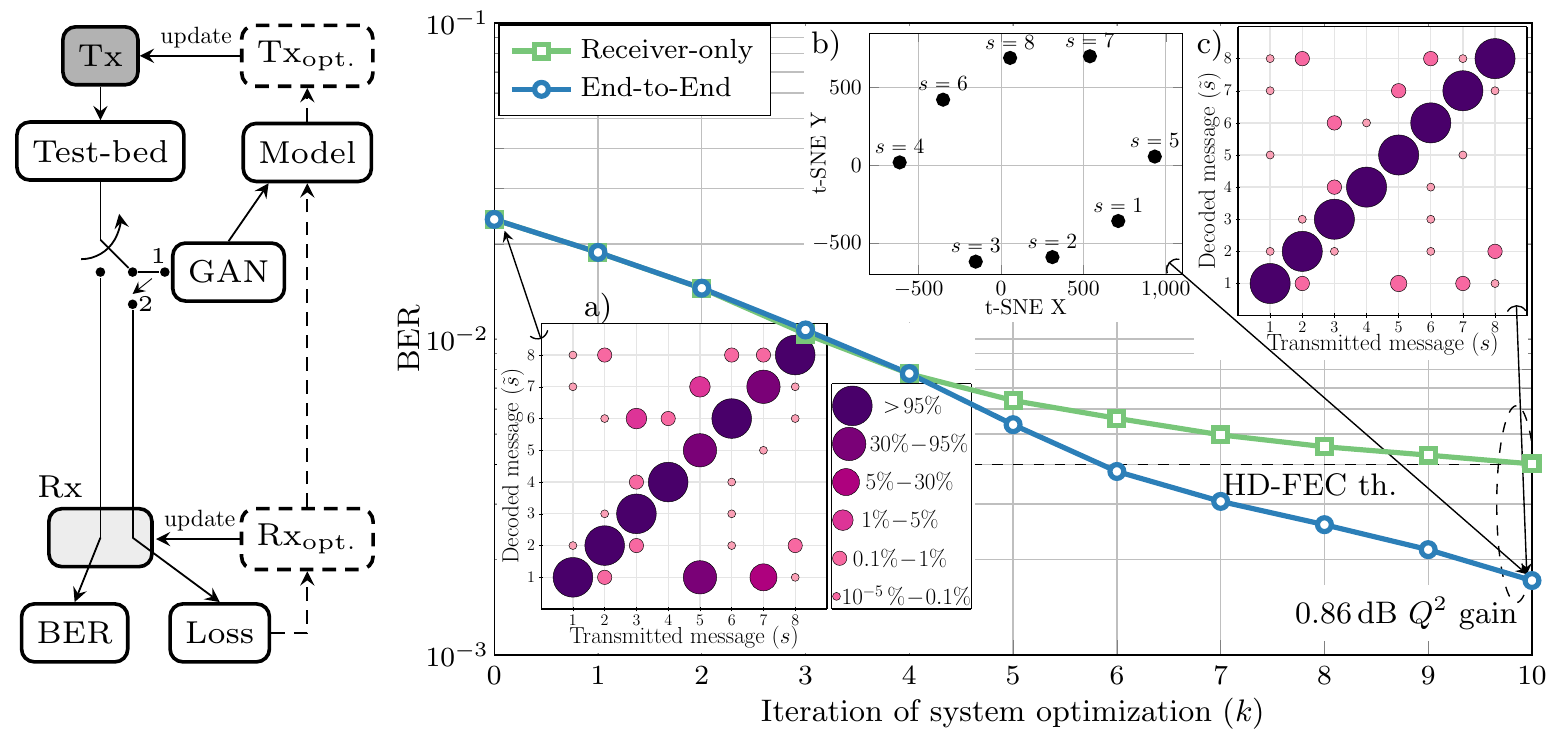}
\vspace*{-5ex}
\caption{(Left) Flow chart of the transmission and optimization algorithm. (Right) Experimental BER as a function of optimization iteration. Insets: a) error probabilities at $k\!=\!0$; b) 2D t-SNE representation of the waveforms output of the transmitter ANN at $k\!=\!10$; c) error probabilities at $k\!=\!10$.}
\label{fig:Exp_results}
\vspace*{-5ex}
\end{figure}
The schematic of the IM/DD test-bed is shown in Fig.~\ref{fig:Exp_setup}, which we describe together with the end-to-end system optimization method, shown in Fig.~\ref{fig:Exp_results}~(left). 
At iteration $k\!=\!0$ the transmitter and receiver ANNs are initialized with parameters trained offline, following~\cite[Sec.~III-D]{Karanov_1}, using the IM/DD channel model of~\cite{Karanov_2}. $N\!\!=\!\!500$ random sequences of $w\!\!=\!\! 8\cdot10^{4}$ messages $\begin{pmatrix}\ldots,s_t,\ldots\end{pmatrix}$, $s_t\!\in\!\{1,\ldots, 8\}$~(3\,bits), are generated and encoded by the transmitter into symbol ($n\!\!=\!\!6$ samples) sequences $\begin{pmatrix}\ldots,\mathbf{x}_t,\ldots\end{pmatrix}$. These are filtered by a 32\,GHz~LPF and applied to a 84\,GSa/s DAC, resulting in a 42\,Gb/s data rate. The waveform is fed to an MZM, biased at the quadrature point, modulating a 1550\,nm laser. The signal is launched at 1\,dBm power into an un-amplifed 20\,km span of SSMF. The power of the received waveform is directly detected by an AC-coupled PIN+TIA and real-time sampled by an ADC. After scaling and offset correction, the digital signal $\begin{pmatrix}\ldots,\mathbf{y}_t,\ldots\end{pmatrix}$ is utilized in two ways: it is fed to the receiver for processing, symbol decision and BER calculation using an optimized bit mapping (see~\cite[Sec.~III-C]{Karanov_3}). Moreover, the transmitted and received sequences are used for GAN training, followed by transceiver optimization. Part of the data ($q\!=\!10^{3}$ elements), grouped as $\mathbf{A}\!:\!=\!\begin{bmatrix}s_1;\ldots;s_{q}\end{bmatrix}\!\in\!{\mathbb{R}}^{q\times1}$ and $\mathbf{B}\!:=\begin{bmatrix}\mathbf{y}_{1};\ldots;\mathbf{y}_{q}\end{bmatrix}\!\in\!{\mathbb{R}}^{q\times n}$, was used for the single step of transceiver learning within this algorithm iteration. The remaining part, assuming $m\!=\!3$, was structured as $\mathbf{C}^{(3)}\!:\!=\![\vv{\mathbf{x}}_{q+2}^{(3)};\ldots;\vv{\mathbf{x}}_{Nw-1}^{(3)}]\!\in\!{\mathbb{R}}^{(Nw-q-2)\times 3n}$, with $\vv{\mathbf{x}}_i^{(3)}\!=\!\begin{pmatrix}\mathbf{x}_{i-1},\mathbf{x}_{i},\mathbf{x}_{i+1}\end{pmatrix}$, $i\!\in\!\{q+2,\ldots,Nw-1\}$, and $\mathbf{D}\!:=\!\begin{bmatrix}\mathbf{y}_{q+2};\ldots;\mathbf{y}_{Nw-1}\end{bmatrix}\!\in\!{\mathbb{R}}^{(Nw-q-2)\times n}$. First, the generative model is trained via the procedure in Sec.~\ref{sec:GAN}. Each row $i$ from $\mathbf{C}^{(3)}$ and $\mathbf{D}$ was used to condition the GAN and provide the ground truth symbol, respectively. Next, the transceiver learning is performed, minimizing the cross entropy $\overline{\mathcal{L}_{\textnormal{syst.}}}(\boldsymbol{\psi}_{\textnormal{Tx/Rx}})=\frac{1}{q}\sum_{i=1}^{q}\ell(s_{i},f_{\textnormal{Rx}}(\mathbf{y}_{i}))$, with $s_i$ and $\mathbf{y}_i$ rows from  $\mathbf{A}$ and $\mathbf{B}$, via SGD using Adam with $10^{-3}$ learning rate. To enable the gradient backpropagation, the generator model is applied in lieu of the test-bed, enabling the update of transceiver parameters~($\psi_{\textnormal{Tx/Rx}}$) in a single process. This completes an iteration of the optimization algorithm. The transmitter and receiver representations are updated and we repeat transmission, GAN training and optimization. GAN training required 500 transmissions of different data sequences, which was the time-limiting step in our setup, as well as stable link parameters to ensure the model validity. Thus, as a proof of concept, we performed 10 optimization iterations, which already resulted in performance gain. The BER of the system during optimization is shown in Fig.~\ref{fig:Exp_results}~(right), with a monotonic reduction observed on each iteration. This improvement is illustrated via insets~a) and~c), showing the error probabilities at $k\!\!=\!\!0$ and $k\!\!=\!\!10$, respectively. No errors exceeding the 1\% threshold occur at $k\!\!=\!\!10$, for which~b) depicts the modulation constellation in 2D via t-SNE dimensionality reduction~\cite{O'Shea_1}. Compared to the previously possible receiver-only optimization on measured data, our method increases the $Q^2$-factor by $0.86$\,dB.\vspace*{-1ex}

\section{Conclusions}\vspace*{-1ex}
We experimentally implemented deep learning of a transceiver based on measured data, an important step towards practical end-to-end optimized transmission. Instead of using an explicit channel model, a generative model of the test-bed was trained to approximate its conditional distribution. It enabled gradient backpropagation in an iteration of the end-to-end system learning. We observed a monotonic decrease in BER on each step of the optimization.

\vspace*{+0.25ex}
\small{\textit{Work under the EU Marie Sk{\l}odowska-Curie project COIN (676448/H2020-MSCA-ITN-2015) \& UK EPSRC TRANSNET.}}

\end{document}